\documentstyle[titlepage]{article}
\newcommand{\R}{{\rm I\kern-2pt R}}
\newtheorem{lemma}{Lemma}[section]
\newtheorem{proposition}{Proposition}[section]

\newtheorem{theorem}{Theorem}[section]
\newtheorem{remark}{Remark}[section]
\newtheorem{definition}{Definition}[section]

\newtheorem{algorithm}{Algorithm}[section]

\begin{document}
\begin{center}
{\bf Algorithms for the Polar Decomposition in Certain Groups and
the Quaternions\\}

Francis Adjei\\
Richland College\\
Dallas, TX 75243 \\
fadjei@dcccd.edu\\

Marcus Cisneros\\
Texas State University\\
San Marcos, TX 78666\\
mdc170@txstate.edu\\

Deep Desai\\
University of Texas at Dallas\\
Richardson, TX 75080\\
dpd170000@utdallas.edu\\

Viswanath Ramakrishna \\
(corresponding author)\\ 
Department of Mathematical Sciences
University of Texas at Dallas\\
Richardson, TX 75080 USA\\
vish@utdallas.edu\\
\& \\
Brandon Whiteley\\
University of Texas at Austin\\
Austin, TX 78705 \\
brandon.whiteley@utexas.edu  
\end{center}

\begin{abstract}
Constructive algorithms, requiring no more than $2\times 2$
matrix manipulations, are provided for finding the entries
of the positive definite
factor in the polar decomposition of matrices 
in sixteen groups preserving a bilinear form
in dimension four. This is used to find quaternionic representations
for these groups analogous
to that for the special orthogonal group. Furthermore, no eigencalculations
are invoked. 
The groups to which the results are applicable include
the symplectic group, and the groups whose signature matrices
are $I_{2,2}$ and $I_{1,3}$ (i.e., the ``Lorentz group").
This is achieved by
characterizing positive definite matrices in these groups.
Together with i) our earlier work on the real symplectic group;
and ii) explicit isomorphisms, obtained by quaternionic  algebra,
between many of these groups, this characterization achieves
the task at hand. A quaint, but key, observation is that 
for symmetric matrices in these groups, positive definiteness
is equivalent to the positive definiteness of diagonal blocks.
For the group whose signature matrix is
$I_{2,2}$ a completion procedure based on this observation
leads to said computation of
the polar decomposition, while for the Lorentz group this is
achieved by passage to its double cover. The latter is aided
by the fact that the inversion of the covering map, when the
target is a positive definite matrix, can be achieved essentially
by inspection as we demonstrate. As byproducts we give a simple
proof of the fact that positive definite matrices in some of
these groups belong to the connected component of the identity, 
find an explicit expression for their logarithm, and
find a simple characterization of the connected components of $G_{I_{2,2}}$
via certain determinants. A characterization of the symmetric
matrices in the connected component of the identity of
these groups in terms of their preimages in the corresponding
covering group is also found.  Finally, quaternionic
representations of elements of these group is found.  

{\it Keywords: Positive definiteness, Schur Complements,
Polar Decomposition, Double Cover, Lorentz Group, Quaternions, Algorithmic}
\end{abstract}

\section{Introduction}
This paper has as its motivation the following seemingly elementary question.
What are the quaternionic representations of certain groups
of real matrices in dimension four, analogous to the representation
of the group $SO(4, \mathbf{R})$ by a pair of unit quaternions?
Given the universal appeal and utility of the quaternions and
its variants, with applications ranging from virtual reality to
photonics and quantum computation 
(see, for instance, \cite{goongi,FourportI,kuiper}), this is a natural
line of investigation.

For other classes of matrices, such as symmetric, antisymmetric and
those in the Lie algebras of the groups alluded to above, expressions
for their quaternionic representations are known \cite{ni,nii,niii}.  
The groups of interest all arise as the isometry groups of non-degenerate
real bilinear forms. In \cite{noncompactportion} this question was
answered for the real symplectic group $Sp(4, \mathbf{R})$.
Note, by the ``symplectic group", reference is being made  
to the real symplectic group, and {\it not the similarly denoted 
$Sp(n)$}, which preserves the standard inner product on $\mathbf{H}^{n}$
(here $\mathbf{H}$ stands for the quaternions). 
The latter is, of course, trivially defined via quaternions. 
In the process of uncovering such a representation a fully constructive
algorithm, not requiring any eigencalculations, for computing the
polar decomposition of a matrix in $Sp(4, \mathbf{R})$ was provided
in \cite{noncompactportion}.

In physical applications where these groups arise, the positive definite
factor in the polar decomposition can be interpreted as the ``lossy"
portion of such matrices, while the orthogonal factor is the conservative
portion. Thus, for instance, if a matrix is in the Lorentz group, the
positive definite factor is precisely the boost factor. Thus, methods
which provide algorithms for the entries of the boost portion, starting
from the matrix being decomposed, are
obviously useful. The Lorentz group (and the related $SO^{}(1,2, \mathbf{R})$)
have recently found many applications in optics (especially
polarization optics), \cite{kim}. It is therefore expected that 
extracting the boost or spatial rotation factor of a Lorentz matrix will
have applications to interpreting the lossy or lossless part of
the corresponding optical element.   

In this work, therefore, we will address algorithms
for the polar decomposition for
fifteen other matrix groups $G_{M}\subseteq M(4, \mathbf{R})$.
Here $G_{M} = \{ X: X^{T}MX = M\}$, with $M$ either one of the sixteen
matrices belonging to the basis for $M(4, \mathbf{R})$,
arising from the isomorphism
of the quaternion tensor product, $\mathbf{H}\otimes
\mathbf{H}$, with $M(4, \mathbf{R})$, or $M= I_{1,3}$
(i.e., the group popularly called the Lorentz group). 
The matrices $M = I_{4}$ and $M= J_{4}$
belong to this basis and the corresponding
cases have already been studied, whence the
count of fifteen.

{\it It is emphasized here that we are not merely providing the structural
form of the factors in the polar decomposition. Rather, we provide
algorithms to determine the entries of the positive definite
factor in the polar
decomposition, starting from the matrix being decomposed, in a fashion
which requires nothing more than $2\times 2$ manipulations.}
Once the positive definite factor has been found, finding the orthogonal
factor is straightforward, since the inverse of a matrix $X$ in these
groups is precisely $\pm MX^{T}M$.
In addition to obtaining the polar decomposition and quaternioinic
representations, the following byproducts also emanate:
i) A simple proof of the fact that positive
definite matrices in $G_{I_{n,n}}$ 
belong to the connected component of the identity [Remark
(\ref{LogofPosDef})], achieved via an {\it explicit} formula for their
logarithm; ii) A characterization of the symmetric matrices
in $SO^{+}(1,3, \mathbf{R})$ and $SO^{+}(2,2,\mathbf{R})$ in terms
of their preimages in the corresponding covering groups [See Theorem
(\ref{PosDefInLorenz}) and Remark (\ref{Whynot})] and iii) An elementary
characterization of the connected components of $G_{I_{n,n}}$
analogous to the well-known such characterization for the Lorentz group,
[see Theorem (\ref{CCCharacterize})].

In the process, we 
prove the following quaint fact which
plays a useful role. A symmetric matrix in one of these
groups $G_{M}$ is positive definite iff its diagonal blocks
are themselves positive definite. As is well known, for a general
real matrix this is only a necessary condition
[see Theorem (\ref{PosDefinso22})].

It is worthwhile to mention a curious trichotomy here. Even though
$Sp(4, \mathbf{R})$ is $10$ dimensional, as opposed to the six dimensional
$G_{I_{2,2}}$ or $G_{I_{1,3}}$, 
the quaternionic description of the former is more elegant
than those of the latter two. In part due to this, one can use the 
quaternionic description of $Sp(4, \mathbf{R})$ itself to obtain the polar
decomposition of matrices in it, whereas for the latter two groups it is
easier to constructively obtain the polar decomposition 
by availing of the block
structure of a matrix in one case and the corresponding
covering spin group in the other. On the other hand obtaining a quaternionic
representation for $G_{I_{2,2}}$, starting from the defining
condition $X^{T}I_{2,2}X = I_{2,2}$, is arduous.
For the group $G_{I_{1,3}}$ the
obstacles corresponding to those for $G_{I_{2,2}}$  become even more
formidable. Therefore, we resort to its double cover $SL(2, \mathbf{C})$.
This is aided by the fact that finding the preimage in $SL(2, \mathbf{C})$
of a positive definite
matrix in the Lorentz group is significantly easier than finding the
preimage of a general element in this group. {\it In fact, it can be performed
essentially by inspection.}  For $SO^{+}(2,2, \mathbf{R})$
the corresponding statement does not hold, whence the need to go the
completion procedure of Sections 3.1 and 3.2,

We conclude this introduction with some history of the linear algebraic
applications of the
isomorphism between $\mathbf{H}\otimes \mathbf{H}$ and $M(4, \mathbf{R})$.
This isomorphism is central to the theory of Clifford algebras, \cite{pertii},
though it is not one of the standard isomorphims of Clifford algebras
with matrix algebras.
However, its usage for 
linear algebraic (especially numerical linear algebraic) purposes is
of relatively recent vintage..
To the best of our knowledge the first instance seems to be the work
of \cite{kyf}, where it was used in the study of linear maps preserving the
Ky-Fan norm. Then in \cite{haconi}, this connection was used
to obtain the Schur canonical form explicitly for real $4\times 4$ 
skew-symmetric matrices. Next, is the work of \cite{nii,ni,niii}, wherein
this connection was put to 
innovative use for solving eigenproblems of
of several classes of structured matrices.
In \cite{expistruc,expisufour}, this isomorphism was used to
explicitly calculate the exponentials of a wide variety 
of $4\times 4$ matrices. In \cite{minpolyi}, explicit formulae for
the minimal polynomials of classes of structured $4\times 4$ matrices
were obtained through this isomorphism. Finally, this isomorphism
played an important role in studying reversion and Clifford conjugation
on a wide variety of Clifford algebras without any intervention of
the corresponding even algebras, \cite{reversion,fpi}.

The balance of this manuscript is organized as follows. In the next section
some notation and useful results on  positive
definite matrices and the algebra isomorphism between 
$\mathbf{H}\otimes \mathbf{H}$ and
$M(4,\mathbf{R})$ are collected. We draw attention to
Remark (\ref{2DSquareroot}), Lemma (\ref{NoDiagEvenfor2by2}),
Remark
(\ref{UsefulHF}) and Remark (\ref{ThreePolarIsEnough}). These play an
important role for what follows.    
The next section elucidates the block structure of positive definite
matrices in the $G_{M}$'s considered here. Section 3.1 shows how to
complete a matrix in $G_{I_{n,n}}$
to positive definiteness given one of its three blocks. Remark
(\ref{LogofPosDef}) is a byproduct of it and is used in Section 5
to characterize the connected components of $G_{I_{2,2}}$.
This completion procedure is important
in Subsection 3.2 which addresses the algorithmic determination of
the positive definite factor in the polar decomposition of a matrix
in $G_{I_{2,2}}$ and also for the quaternionic representation
of matrices in $G_{I_{2,2}}$, described in
Section 4.
Section 5 studies positive definiteness
in the Lorentz group, showing among other things that the only other symmetric
matrices in the Lorentz group are all indefinite - see
Theorem (\ref{PosDefInLorenz}). A byproduct of the proof of
this theorem is Algorithm (\ref{PosDefInversioninLorentz})
which shows that the preimage of a positive definite matrix in
$SL(2, \mathbf{C})$, under the covering map, can be found by inspection.
We also draw attention
to Remark (\ref{Whynot}) which explains how much the analogous
calculation can be carried out for $SO^{+}(2,2,\mathbf{R})$.
Theorem (\ref{CCCharacterize}) is a byproduct of Remark (\ref{Whynot}).
It presents a characterization of the connected components
of $G_{I_{2,2}}$. 
 The final section offers conclusions.
An appendix provides the statement of our result from \cite{noncompactportion}
on the quaternionic representation of symplectic, positive definite
matrices for the purposes of comparison. 
  
\section{Notation and Preliminary Observations}
The following definitions, notations and results will 
be frequently met in this work:
\begin{itemize} 
\item $M(4,\mathbf{R})$ (also denoted $gl (4, \mathbf{R})$) 
is the algebra of real
$4\times 4$ matrices.

\item Let $M$ be an $n\times n$ real invertible matrix.
Then $G_{M} = \{X\in M(n,\mathbf{R}): X^{T}MX = M\}$.
$G_{M}$ is a Lie group. There are obvious extensions to the complex
case (both with transposition and Hermitian conjugation in the
definition), but for the immediate purposes of this work, this
definition of $G_{M}$ suffices.
\item 
$I_{p,q} = \left ( \begin{array}{cc}
I_{p} & 0_{p\times q}\\
0_{q\times p} & I_{q}
\end{array}
\right )$.
\item $J_{2n} = \left ( \begin{array}{cc}
0_{n} & I_{n}\\
-I_{n} & 0_{n}
\end{array}
\right )$. $Sp(2n, R)$ is the standard notation
for $G_{J_{2n}}$.

\item We use the standard notatation $SO(p,q, \mathbf{R})$ for the
determinant $1$ matrices in $G_{I_{p,q}}$ and $SO^{+}(p, q, \mathbf{R})$
for the connected component of the identity in $SO(p,q, \mathbf{R})$.
The Lorentz group is then $SO^{+}(1, 3, \mathbf{R})$.

\item Essential use will be made of the following theorem
(with statement adapted to the needs of this work,
see \cite{structureddecompI}):
\begin{proposition}
\label{polardecompi}
{\rm Let $X \in G_{M}$, where $M = cU$, with $c>0$ and
$U$ real orthogonal. If $X = QP$ is its
polar decomposition, with $P$ positive definite and $Q$
real orthogonal, then $P$ and $Q$ are also in $G_{M}$.}
\end{proposition}
\end{itemize}
We next collect some definitions and results on real positive 
definite matrices. Most statements 
may be found in \cite{hhorni}.
\begin{definition}
\label{sqrts}
{\rm Let $Y$ be a real positive definite matrix. A real square matrix
$Z$ satisfying $Y = Z^{T}Z$ is said to said to be a square root of $Y$.}
\end{definition}

\begin{remark}
\label{2DSquareroot}
{\rm Square roots of positive definite matrices are not unique. However,
if $Z_{1}$ is a square root of $Y$ then $Z_{2}$ is also a square root
of $Y$ iff there exists a real orthogonal matrix $U$ such that $Z_{2}
= UZ_{1}$. For $2\times 2$ matrices, therefore, 
\begin{equation}
\label{UTheta}
U = U_{\theta}
= \left (\begin{array}{cc}
\cos \theta & -\sin \theta\\
\sin \theta & \cos \theta
\end{array}
\right )
\end{equation}
if the determinants of $Z_{1}$ and $Z_{2}$ share the same sign,
whereas, 
\begin{equation}
\label{VTheta}
U = V_{\theta} = 
\left ( \begin{array}{cc}
\cos \theta & \sin \theta\\
\sin \theta & -\cos \theta
\end{array}
\right )
\end{equation}
if the determinants of $Z_{1}$ and $Z_{2}$ have opposite signs.}
\end{remark}

Among these square roots, there is a unique upper triangular
one with positive diagonal
entries (that provided by the Cholesky factorization), say $Z_{1}$ and
a unique positive definite one, say $Z_{2}$. Thus, there is a
real special orthogonal $U$ such that $X_{2} = UX_{1}$.

The following lemma shows, by way of variety, how to find $X_{2}$ in closed
form for $2\times 2$
matrices 
without any recourse to 
any eigencalculations, though it can easily be computed in closed
form by an orthogonal diagonalization. 

\begin{lemma}
\label{NoDiagEvenfor2by2}
{\rm If $Y = \left (\begin{array}{cc}
	a_{11} & a_{12}\\
	a_{12} & a_{22}
\end{array}
\right )$ is positive definite, 
then its unique positive definite
square root is $X_{2} = \left (\begin{array}{cc}
\alpha\cos\theta & \alpha \sin\theta\\
\alpha\sin\theta & \beta \sin\theta + \gamma\cos\theta
\end{array}
\right )$, where $\alpha = \sqrt{a_{11}}, \beta = 
\frac{a_{12}}{\sqrt{a_{11}}}; \gamma = \sqrt{\frac{{\mbox det}(Y)}{a_{11}}}$
and $\tan\theta = \frac{\beta}{\alpha + \gamma}$, where $\theta$ is chosen
to be in the first quadrant if $a_{12} > 0$ and in the fourth quadrant
if $a_{12} < 0$. If $a_{12} = 0$, then $\theta = 0$.}
\end{lemma}

\noindent {\bf Proof:}
The upper Cholesky factor of $Y$ is $X_{1}
= \left (\begin{array}{cc}
 \alpha & \beta\\
0  & \gamma 
\end{array}
\right )$. Therefore, since the determinants of $X_{1}$ and $X_{2}$ are
both positive, $X_{2} = U_{\theta}X_{1}$, with $U_{\theta}$ as in
Remark (\ref{2DSquareroot}). The $(1,2)$ entry of $UX_{1}$ is
$\beta\cos\theta - \gamma\sin\theta$, while the $(2,1)$ entry is
$\alpha\sin\theta$. Since $UX_{1}$ has to be symmetric, we find
$\tan\theta = \frac{\beta}{\alpha + \gamma}$. Next, since ${\mbox det}
(X_{2}) = {\mbox det}
(X_{1})$, we need the $(1,1)$ entry of $UX_{1}$ to be positive. But this
entry is $\alpha\cos\theta$. Since $\alpha > 0$, we need
$\cos\theta > 0$. Since only $\beta$ can be negative if at all
(which happens precisely when $a_{12} < 0$), we obtain the quadrant
characterization of $\theta$ in the statement. $\diamondsuit$.

Use of the following remark will be made in Sec 3.2.

\begin{remark}
\label{AMinusI}
{\rm Let $A > 0$ and $A^{2}-I\geq 0$. Then $A-I\geq 0$ also.
This is elementary, since $\lambda > 0$ and $\lambda^{2}-1\geq 0$
together imply $\lambda - 1 \geq 0$, where $\lambda$ is an eigenvalue
of $A$.}
\end{remark}

\noindent Next relevant definitions and results regarding
quaternions and their connection to real matrices will be presented.
Throughout $\mathbf{H}$ will be denote the skew-field (
the division algebra) of the {\it quaternions},
while $\mathbf{P}$ stands for the
{\it purely imaginary} quaternions, tacitly identified with $R^{3}$.

\noindent {\bf $\mathbf{H}\otimes \mathbf{H}$
and $M_{4}(\mathbf{R})$}: The algebra isomorphism
between $\mathbf{H}\otimes \mathbf{H}$ 
and $M_{4}(\mathbf{R})$, which is used here is
the following:

\begin{itemize}
\item Associate to each product
tensor $p\otimes q\in \mathbf{H}\otimes \mathbf{H}$ 
the matrix, $M_{p\otimes q}$, of the map which sends $x\in \mathbf{H}$
to $px\bar{q}$, identifying $R^{4}$ with $\mathbf{H}$ via the 
basis $\{1,i,j,k\}$.
Here, $\bar{q}$ is the conjugate of $q$. 
This yields an algebra isomorphism
between $\mathbf{H}\otimes \mathbf{H}$ 
and $M_{4}(\mathbf{R})$.  
\item Define conjugation in $\mathbf{H}\otimes \mathbf{H}$ by first defining
the conjugate 
of a decomposable tensor $a\otimes b$ as $\bar{a}\otimes \bar{b}$, and then   
extending this to all
of $\mathbf{H}\otimes \mathbf{H}$ by linearity. 
Then $M_{\bar{a}\otimes\bar{b}}
= (M_{a\otimes b})^{T}$.
Thus, the most general element of $M_{4}(\mathbf{R})$ admits the quaternion
representation $a1\otimes 1 + p\otimes i + q\otimes j + r\otimes k
+ s\otimes 1 + 1\otimes t$, with $a\in \mathbf{R}$ and $p,q,r,s,t 
\in \mathbf{P}$.
The summand $a1\otimes 1 + p\otimes i + q\otimes j + r\otimes k$
is the symmetric part of the  matrix, while the summand
$ s\otimes 1 + 1\otimes t$ is the skew-symmetric part of the matrix.
Expressions for $a,p,q,r,s,t$ (which are linear in the entries of
the matrix being represented) are easy to find, \cite{ni}. Finally
$4a$ is the trace of the matrix.  
\end{itemize}

\begin{remark}
\label{UsefulHF}
{\rm Several useful facts concerning the aforementioned isomorphism
are collected below.
\begin{enumerate}
\item The sixteen matrices $M_{e\otimes f}$, where $e, f \in
\{1,i, j, k\}$ form a basis, denoted $\mathbf{B}$,
of special orthogonal matrices for
$M(4, \mathbf{R})$. Of these $10$ are symmetric ($M_{1\otimes 1}$ and
those for which neither $e$ nor $f$ equals $1$) and the remaining
six are antisymmetric.

\item Of special relevance to this work are $M_{i\otimes i}$ which
equals $I_{2,2}$ and $M_{1\otimes j}$ which equals $J_{4}$.
Other useful members of this basis are $M_{i\otimes j}$, the
$4\times 4$ flip matrix; $M_{i\otimes k}$ and $M_{i\otimes 1}$.
In \cite{reversion,fpi} essential use of $M_{1\otimes i},
M_{1\otimes k}$ and $M_{k\otimes 1}$ was made to study algorithmically the
spin groups.
  
\item One can find explicit congruences (simultaneously similarities)
between the fifteen symmetric members of this basis, which do not
equal $I_{4}$. This task can be reduced to elementary algebraic
calculations in $\mathbf{H}$. For instance,
to find a similarity between $M_{i\otimes i}$ and $M_{i\otimes j}$,
we find unit quaternions $p, q$ satisfying $\bar{p}ip = i$ and
$\bar{q}iq = j$. One choice is $p= \frac{1}{\sqrt{2}} (j + k)$
and $q = \frac{1}{\sqrt{2}} (i + j)$.

\item Similarly, we can find explicit special orthogonal similarities between
any two of $M_{1\otimes i}, M_{1\otimes j}, M_{1\otimes k}$. For instance,
we find a unit quaternion $q$ satisfying $\bar{q}jq = i$. Then
$S^{-1}M_{1\otimes j}S = M_{1\otimes i}$ where $S = M_{1\otimes q}$.
One choice is $q = \frac{1}{\sqrt{2}} (i + j)$ again.
Similarly, we can find explicit special orthogonal similarities between
any two of $M_{i\otimes 1}, M_{j\otimes 1}, M_{k\otimes 1}$.
Next notice that $M_{i\otimes 1} = - {\mbox diag} (J_{2}, J_{2})$ and
$J_{4} = M_{1\otimes j}$. Since,
$P^{T}J_{4}P = {\mbox diag} (J_{2}, J_{2})$, where $P$ is the permutation
matrix which, in column form equals 
\[
[e_{1}\mid e_{3}\mid e_{2}\mid e_{4}]
\] it is seen that one can find explicit
similarities between any two 
of \[
\{M_{1\otimes i}, M_{1\otimes j}, M_{1\otimes k},
-M_{i\otimes 1}, -M_{j\otimes 1}, -M_{k\otimes 1}\}.
\]
\end{enumerate}
}

\end{remark}
\begin{remark}
\label{ThreePolarIsEnough}

{\rm Observing that $G_{-M} = G_{M}$ and that if $M_{2} = S^{T}M_{1}S$
for an explicit orthogonal $S$ (as supplied by the previous two items
of the previous remark),
then $X\in G_{M_{2}}$ iff $SXS^{T}\in G_{M_{1}}$, we see that it
suffices to explicitly compute the polar decomposition of matrices
in $G_{I_{4}} = O(4, \mathbf{R})$, $G_{I_{2,2}}$ and $G_{J_{4}} =
Sp(4, \mathbf{R})$ to be able to compute the polar decomposition
explicitly in any of the sixteen $G_{M}$'s, for $M\in \mathbf{B}$.
Indeed, if $X\in G_{M_{2}}$ and if the polar decomposition 
of $SXS^{T}\in G_{M_{1}}$ is $SXS^{T} = QP$, then as the matrices
$S^{T}QS$ and $S^{T}PS$ are orthogonal and positive definite
respectively, it follows from uniqueness in the polar decomposition
for invertible matrices that $X = (S^{T}QS)(S^{T}PS)$ is the
polar decomposition of $X$.    
		
For the polar decompsition of $O(4, \mathbf{R})$,
there is nothing to do, while for $Sp(4, \mathbf{R})$ this task was performed
in \cite{noncompactportion}. Therefore, it remains to do the needful for
the groups $G_{I_{2,2}}$ and $G_{I_{1,3}}$ .}
\end{remark}

\section{Positive Definiteness in Certain Groups}

We begin with a characterization of symmetric matrices in
$G_{M}$, when $M= I_{p,q}$.

\begin{lemma}
\label{symmetricinso22}
{\rm Let $X = \left (\begin{array}{cc}
A & B\\
B^{T} & D
\end{array}
\right )$ be a symmetric $n\times n$ real matrix. Let $A$ be $p\times p$
and $D$ $q\times q$ with $p+ q = n$. Then $X^{T}I_{p,q}X
= I_{p,q}$ iff the following hold:
\begin{enumerate}
\item $A^{2} - BB^{T} = I_{p}$.
\item $AB = BD$
\item $D^{2} - B^{T}B = I_{q}$.
\end{enumerate}
}
\end{lemma}
\noindent {\bf Proof:} This is a straightforward block matrix calculation.

Next we will characterize when an $X$, as in Lemma 
(\ref{symmetricinso22}) is positive definite. Clearly a necessary condition
is that the diagonal blocks themselves are positive definite. For general
$X$ this is not even remotely sufficient. But if $X$ is also
in $G_{M}$, with $M=I_{p,q}$, then it is sufficient as the following result
shows.

\begin{theorem}
\label{PosDefinso22} 
{\rm A symmetric $n\times n$ real symmetric matrix, satisfying
$X^{T}I_{p,q}X = I_{p,q}$ is positive definite iff its diagonal
square blocks are both positive definite.}

\end{theorem}

\noindent {\bf Proof:} The necessity is well known. Conversely, let $A  > 0$
and $D > 0$. We need to show that the Schur complement
$\Delta_{A} = D - B^{T}A^{-1}B > 0$ also. From 2) of Lemma
(\ref{symmetricinso22}) we get $AB = BD$ and hence $B = A^{-1}BD$.
So $B$ intertwines $A^{-1}$ and $D^{-1}$ also.

Next, 3) of
Lemma (\ref{symmetricinso22}) gives $D^{2} = I_{q} + BB^{T}$.  Hence,
$D = D^{-1} + B^{T}BD^{-1}$. Hence, $D = D^{-1} + B^{T}A^{-1}B$.
Therefore, $\Delta_{A} = D - B^{T}A^{-1}B = D^{-1} + B^{T}A^{-1}B
- B^{T}A^{-1}B = D^{-1}$.
As $D^{-1} > 0$ also, we obtain the desired conclusion. $\diamondsuit$

\begin{remark}
{\rm One can also show that $\Delta_{D}$, the Schur complement of $D$,
is $A^{-1}$.}
\end{remark}

\begin{remark}
\label{LorenzToo}
{\rm There are interesting analogues of Lemma (\ref{symmetricinso22}) and
Theorem (\ref{PosDefinso22}) for other groups. 
For instance,
\begin{enumerate}
\item Though the group $G_{I_{1,3}}$ has been subsumed above, it is convenient
to view matrices in it as $2\times 2$ block matrices and the signature
matrix $I_{1,3}$ as ${\mbox diag}(\sigma_{z}, - I_{2})$. If one does this
a symmetric matrix is in $G_{I_{1,3}}$ iff
\begin{equation}
\label{SymmetricinLorenz}
A\sigma_{z}A - BB^{T} = \sigma_{z}, A\sigma_{z}B = BD;
D^{2} - B^{T}\sigma_{z}B = I_{2}
\end{equation}
Using this one shows again that the Schur complement of $A$ is $D^{-1}$.
\item  Similarly, let $X$ be $2n\times 2n$ real symmetric. Then $X$
is real symplectic iff $BA$ and $DB$ are symmetric and $AD = I_{n} + B^{2}$.
Once again such an $X$ is positive definite iff its diagonal blocks
are themselves positive definite. For instance, the Schur complement
of $A$ is $D - B^{T}A^{-1}B
= D + A^{-1}BAA^{-1}B = D + A^{-1}B^{2} = D + A^{-1}(AD - I) = A^{-1}$.

\item Consider the $4\times 4$ flip matrix
$F_{4}$ (which is $M_{j\otimes i})$. Its higher dimensional analogue
is $F_{2n}$. Consider a $2n\times 2n$ symmetric matrix
$X = \left (\begin{array}{cc}
A & B\\
B^{T} & D
\end{array}
\right )$. Then $X^{T}F_{2n}X = F_{2n}$ iff we have the following
conditions: i) $BF_{n}A$ and $DF_{n}B$ are antisymmetric and
$BF_{n}B + AF_{n}D = F_{n}$.

Let $A > 0$. We will show that $\Delta_{A} > 0$ by showing that
$F_{n}^{T}\Delta_{A}F_{n} > 0$. Now $F_{n}^{T}\Delta_{A}F_{n}
= F_{n}\Delta_{A}F_{n} = F_{n} (D - B^{T}A^{-1}B)F_{n}$.
But from $BF_{n}A = -AF_{n}B^{T}$, we get $BF_{n} = - AF_{n}B^{T}A^{-1}$ and
hence $BF_{n}BF_{n} = -AF_{n}B^{T}A^{-1}F_{n}$.

On the other hand, $BF_{n}B = F_{} - AF_{n}D$, whence $A^{-1}BF_{n}BF_{n}
= A^{-1} (F_{n}^{2} - A^F_{n}D)F_{n} = A^{-1} - F_{n}DF_{n}$, since
$F_{n}^{2} = I_{n}$. Therefore, $F_{n}\Delta_{A}F_{n} = A^{-1} > 0$.

\item  Let  $M= M_{i\otimes k} = 
\left (\begin{array}{cc} 
0_{2} & I_{2}\\
I_{2} & 0_{2}
\end{array}
\right )$. Its $2n$ dimensional generalization is
$K_{2n} = \left (\begin{array}{cc} 
0_{n} & I_{n}\\
I_{n} & 0_{n}
\end{array}
\right )$. Now $X$ is symmetric and in $G_{M}$
iff $BA$ and $DB$ antisymmetric and $B^{2} = -AD$, as a direct calculation
shows. As before for $X$ to be positive definite we need
$A > 0$ and $D> 0$. These conditions are also sufficient.
For instance, the Schur complement of $A$ is $D - B^{T}A^{-1}B
= D + A^{-1}BAA^{-1}B = D + A^{-1}B^{2} = D + A^{-1}(I-AD) = A^{-1}$.
Here in the first equality we have used $B^{T} = -A^{-1}BA$ and
in the penultimate equality $B^{2} = I-AD$ has been employed.
 
\item In general the sixteen $M$'s arising from
the $\mathbf{H}\otimes \mathbf{H}$ basis are all either 
block diagonal or block reverse diagonal.
For symmetric $X\in G_{M}$, for the former class the pattern is that
$\Delta_{A}$ is congruent to $D^{-1}$, while for the latter it is congruent
to $A^{-1}$.
  
\end{enumerate}
}
\end{remark}   

\subsection{Completion to Positivity in $G_{I_{n, n}}$}
Let us now discuss how one may ``complete" $A$, $D$ or
$B$ to obtain a positive definite matrix in $G_{I_{n,n}}$. The constructions
in this section are {\bf crucial} for finding the polar decomposition
of a matrix in $G_{I_{2,2}}$ to be discussed in Section 3.2. In addition,
they provide a constructive proof of the fact that positive definite
matrices in $G_{I_{n, n}}$ belong to $SO^{+}(n, n, \mathbf{R})$. They also
provide the quaternionic representation of a positive definite
matrix in $G_{I_{2,2}}$. 
\begin{itemize}
\item {\it Completion given $A$:}
Suppose $A > 0$ and $A^{2} - I_{n}\geq 0$. We first pick any square root
$B^{T}$ of $A^{2} - I_{n}$, i.e., a $n\times n$ $B$ satisfying
$A^{2} - I = BB^{T}$. Next let $D$ be the unique positive definite square
root of $I_{n} + B^{T}B$. Clearly then the data satisfy requirements
$1$ and $3$ of Lemma (\ref{symmetricinso22}). The claim is that this
data also satisfy requirement $2$ of Lemma (\ref{symmetricinso22}).
If this is demonstrated, $X = \left (\begin{array}{cc}
A & B\\
B^{T} & D
\end{array}
\right )$ will be positive definite in light of 
Theorem (\ref{PosDefinso22}).

Towards that end note that $A^{2}B = (I + BB^{T})B = B + BB^{T}B$,
while $BD^{2} = B(I + B^{T}B) = B + BB^{T}B$. Thus, the $B$ constructed 
above automatically interwines $A^{2}$ and $B^{2}$. 

Let $U, V$ be
real orthogonal matrices which diagonalize
$A$ and $D$ respectively. Then from $A^{2}B = BD^{2}$, we get
\begin{equation}
\label{FC1}
(U^{T}A^{2}U) (U^{T}BV) = (U^{T}BV)(V^{T}D^{2}V)
\end{equation} Let $\lambda_{i},
\mu_{j}$ be the eigenvalues of $A$ and $D$ respectively.
Then Equation (\ref{FC1}) is equivalent to
\begin{equation}
\label{FC2}
(\lambda_{i} ^{2} - \mu_{j}^{2}) (U^{T}BV)_{ij} = 0
\end{equation}

As $\lambda_{i} + \mu_{j} > 0$, we then obtain that 
$(\lambda_{i} - \mu_{j}) (U^{T}BV)_{ij} = 0$. This is, of course,
equivalent to $(U^{T}AU)(U^{T}BV) = (U^{T}BV))(V^{T}DV)$, which in turn
is equivalent to $AB = BD$.

\item  {\it Completion given $D$:}
Let $D > 0$ and $D^{2} - I\geq 0$. First pick $B$ to be any $n\times n$ 
square root of $D^{2} - I_{n}$. Then pick $A$ to be the unique
positive definite square root of $I+ BB^{T}$. Then, as before 
$A^{2}B = BD^{2}$. Therefore, just as in the item above,
$AB = BD$.        

\item {\it Completion given $B$:} First pick $A$ to be the unique positive
definite square root of $I + BB^{T}$ and then $D$ to be the unique positive
definite square root of $I + B^{T}B$. Then a calculation shows that
$A^{2}B = BD^{2}$. Therefore, as in the item above,
we also get $AB = BD$.  
\end{itemize}

\begin{remark}
\label{LogofPosDef}
{\rm It is perhaps folklore that the positive definite matrices
in $G_{I_{p,q}}$ belong to $SO^{+}(p, q, \mathbf{R})$. Here we give
a simple constructive proof, which also provides an explicit formula
for ``the" logarithm of such a matrix for the case of interest to
us, namely $G_{I_{n,n}}$. Let $P$ be such a positive definite
matrix. To show that it belongs to $SO^{+}(n, n, \mathbf{R})$, it
suffices to express it as the exponential of an element
in the Lie algebra of $SO^{+}(n, n, \mathbf{R})$. Since ``the"
logarithm of a positive definite matrix is also symmetric.
We seek a matrix of the form $\left(\begin{array}{cc}
0 & X\\
X^{T} & 0
\end{array}
\right )$. Let the $(1,2)$ block of $P$ be $B$. By the discussion
on completion given $B$, it is necessary and sufficient
to find an $X$ so that the $(1,2)$ block of
${\mbox exp} [\left(\begin{array}{cc}
0 & X\\
X^{T} & 0
\end{array}
\right ) ]$ equals $B$. Let $B = UDV^{T}$ be the singular value decomposition
of $B$, with $\sigma_{i}$ the singular values of $B$. 
We let $X = UEV^{T}$, where $E$ is a diagonal matrix of
non-negative entries to be chosen. With this choice of $X$, a calculation
which is omitted shows that if we pick $E = {\mbox diag}(y_{1}, \ldots ,
y_{n})$, where $y_{i}= \ln [\frac{\sigma_{i} + \sqrt{\sigma_{i}^{2} + 1}}{2}]$,
then indeed the $(1,2)$ block of ${\mbox exp} [\left(\begin{array}{cc}
0 & X\\
X^{T} & 0
\end{array}
\right ) ]$ is $B$ and hence that $P$ is the exponential of
$\left(\begin{array}{cc}
0 & X\\
X^{T} & 0
\end{array}
\right )$. Thus, $P\in SO^{+}(n,n, \mathbf{R})$.}
\end{remark}

Remark (\ref{LogofPosDef}) will be used in Theorem (\ref{CCCharacterize})
later to characterize the connected components of $G_{I_{2,2}}$.

\subsection{Computing the polar decomposition in $G_{I_{2,2}}$:}

We now specialize to $n=2$. Then the constructions below lead to
the positive definite factor in the polar decomposition of a matrix
$X_{4\times 4}$ satisfying $X^{T}I_{2,2}X = I_{2,2}$ without requiring even 
$2\times 2$ eigencalculations. These manipulations carry over to
$G_{I_{n,n}}$ or more generally $G_{I_{p,q}}$, except that 
we will need eigencalculations for $n > 2$,
which are typically no longer possible in closed form.

So let $Z$  satisfy $Z^{T}I_{2,2}Z = I_{2,2}$. Then $X = Z^{T}Z$ is both
positive definite and satisfies $X^{T}I_{2,2}X = I_{2,2}$. Therefore,
$Y$, the unique positive definite square root also
satisfies $Y^{T}I_{2,2}Y = I_{2,2}$.

Suppose \[
X = \left (\begin{array}{cc}
A& B\\
B^{T} & D
\end{array}
\right )
\]
Thus, $A, B$ and $D$ are known a priori  and satisfy the conditions in
Lemma (\ref{symmetricinso22}) and
Theorem (\ref{PosDefinso22}). In particular $A$ and $D$ are positive definite
with $A^{2} - I$ and $D^{2} -I$ both positive semidefinite.
In light of Remark (\ref{AMinusI})
it then follows that $A - I\geq 0$.

We then have to find
\[
P = \left (\begin{array}{cc}
E & F\\
F^{T} & H
\end{array}
\right )
\]
with i) $P^{2} = X$; ii) $E, F, H$ satisfying the conditions
in Lemma ( \ref{symmetricinso22}) and Theorem (\ref{PosDefinso22}).

The condition $P^{2} = X$ is equivalent to i) $A = E^{2} + FF^{T}$;
ii) $B = EF + FH$ and iii) $D = F^{T}F + H^{2}$.

Since necessarily $E^{2} - I = FF^{T}$, we find that necessarily
$A = I + (\sqrt{2}F)(\sqrt{2}F)^{T}$.
Since, $A - I\geq 0$, we can find by inspection its lower Cholesky
factor $G$. 
Then the most general square root is $G_{\theta}^{T}$, where
$G_{\theta} = G R_{\theta}$ with $R_{\theta}$ is either
$U_{\theta}$ or $V_{\theta}$ as given by Equations (\ref{UTheta}) or
(\ref{VTheta}) respectively.

Let $F_{\theta_{i} } = \frac{1}{\sqrt{2}}G_{\theta_{i} }$,
where $\theta_{1}$ corresponds to the choice of $U_{\theta}$ and
$\theta_{2}$ to that of $V_{\theta}$.
Correspondingly let $E_{\theta_{i} }, H_{\theta_{i}}$ be 
the unique positive definite
square  roots of $I + F_{\theta_{i}} F_{\theta_{i}}^{T}$ and $I +
F_{\theta_{i} }^{T}F_{\theta_{i}}$ respectively (these can be obtained 
in closed form, e.g., via Lemma (\ref{NoDiagEvenfor2by2}).
Note, thus that $E_{\theta_{i}}$ is independent
of $\theta$  while $H_{\theta_{i}}$ is a  
$2\times 2$ matrix 
whose entries are some explicitly computable trigonometric
functions of $\theta_{i}$.

By the Section 3.1, therefore the matrices 
$
\left (\begin{array}{cc}
E_{\theta_{i}} & F_{\theta_{i}}\\
F_{\theta_{i}}^{T} & H_{\theta_{i}}
\end{array}
\right )$ are guaranteed to be positive definite and in
$G_{I_{2,2}}$. In particular, $E_{\theta_{i}}F_{\theta_{i}}
= F_{\theta_{i}}H_{\theta_{i}}, i=1,2$.

Suppose at least one of the $\theta_{i}$ can be
chosen to satisfy $E_{\theta_{i} }
F_{\theta_{i}} +
F_{\theta_{i}}H_{\theta_{i}} = B$ (we will see presently that
precisely one of the $\theta_{i}$ can be so picked). Then we claim
that the
remaining two conditions, viz., $A = E_{\theta_{i}}^{2} + 
F_{\theta_{i}}F_{\theta_{i}}^{T}$;
and $D = F_{\theta_{i}}^{T}F_{\theta_{i}} + H_{\theta_{i}}^{2}$
will be satisfied. 

For instance, to show the latter we proceed as follows.
First observe that both $D$ and 
$F_{\theta_{i}}^{T}F_{\theta_{i}} + H_{\theta_{i}}^{2}$ are positive
definite matrices (the latter because $H_{\theta_{i}} > 0$).
So, as $D$ is the unique positive definite square
root of $I + B^{T}B$, 
it suffices to demonstrate that 
\begin{equation}
\label{Life}  
(F_{\theta_{i}}^{T}F_{\theta_{i}} + H_{\theta_{i}}^{2})^{2}
= I + B^{T}B
\end{equation} 

Now a calculation, which repeatedly uses $H_{\theta_{i}}^{2} = 
I + F_{\theta_{i}}^{T}F_{\theta_{i}}$ and 
$E_{\theta_{i}}F_{\theta_{i}} = F_{\theta_{i}}H_{\theta_{i}}$,
	confirms Equation (\ref{Life}).
Similarly, $A = E_{\theta_{i}}^{2} + F_{\theta_{i}}
F_{\theta_{i}}^{T}$.

Now the equation $E_{\theta_{i} }
F_{\theta_{i}} +
F_{\theta_{i}}H_{\theta_{i}} = B$ is, of course,
\[
2E_{\theta{i}}F_{\theta_{i}} =B 
\]
which, in turn, is
\begin{equation}
\sqrt{2} (I_{2} + \frac{1}{2}GG^{T})^{1/2}GR_{\theta_{i}}
= B 
\end{equation}

Now this last equation  has at least one solution, corresponding to
either $R_{\theta_{1}} = U_{\theta_{1}}$ or to $R_{\theta_{2}}
= V_{\theta_{2}}$ 
because there is 
precisely one positive definite square root of $X^{T}X$ and this is guaranteed
to be in $G_{I_{2,2}}$. In fact, if ${\mbox det}(B) > 0$, then there is
a solution $\theta_{1}$, while if ${\mbox det}(B) < 0$, there is a solution
$\theta_{2}$. If ${\mbox det}(B) = 0$, then both equations may have to be 
addressed, with either both possessing
a solution or  one possessing a solution and the other none. In the case
of multiple solutions, we are guaranteed that the corresponding 
$G_{\theta}$ is unique, because of the uniqueness of the 
positive definite square root.

Let this value of $\theta_{1}$ (or $\theta_{2}$,
as the case may be) be denoted $\theta_{0}$. Then      
defining $F = F_{\theta_{0}}$, $E= E_{\theta_{0}}$ and
$H_{\theta_{0}}$, it follows that
\[
P_{\theta_{0}} =
\left (\begin{array}{cc}
E_{\theta_{0}} & F_{\theta_{0}}\\
F_{\theta_{0}}^{T} & H_{\theta_{0}}
\end{array}
\right )
\] is the positive definite factor in the polar decomposition
of $X$.

Let us summarize the paragraphs above as an algorithm:

\begin{algorithm}
\label{Compute Polar}
{\rm Given $X\in G_{I_{2,2}}$, the following algorithm
computes the positive definite factor, $P$, in the polar decomposition
$X = PQ$.
\begin{enumerate}
\item Compute $X^{T}X = \left (\begin{array}{cc}
A & B\\
B^{T} & D
\end{array} \right )$. Thus, the blocks $A, B, D$ satisfy the
conditions in Theorem (\ref{PosDefinso22}). In particular,
$A - I_{2}\geq 0$.
\item Let $G$ be the lower triangular factor in the Cholesky
factorization of $A-I$. Let $U_{\theta_{1}}$ be as in Equation
(\ref{UTheta}) and $V_{\theta_{2}}$ be as in Equation (\ref{VTheta}).
\item If ${\mbox det}(B) > 0$, then find $\theta_{1}$
from the equation 
$\sqrt{2} (I_{2} + \frac{1}{2}GG^{T})^{1/2}GU_{\theta_{1}}
= B$. If ${\mbox det}(B) < 0$, then find $\theta_{2}$
from the equation $\sqrt{2} (I_{2} + \frac{1}{2}GG^{T})^{1/2}GV_{\theta_{2}}
= B$. If ${\mbox det}(B) = 0$ then at least one of the two equations
$\sqrt{2} (I_{2} + \frac{1}{2}GG^{T})^{1/2}GU_{\theta_{1}}
= B$ or $\sqrt{2} (I_{2} + \frac{1}{2}GG^{T})^{1/2}GV_{\theta_{2}}
= B$  will have a solution.  
Let the solution, $\theta_{1}$ or $\theta_{2}$, be denoted by
$\theta_{0}$.
\item 
$F_{\theta_{0}} = \frac{1}{\sqrt{2}}GR_{\theta_{0}}$,
where $R_{\theta_{0}}$ is $U_{\theta_{1}}$ if $\theta_{0}
= \theta_{1}$ or $V_{\theta_{2}}$ if $\theta_{0} =\theta_{2}$.
Let $E_{\theta_{0}}$
be the unique positive definite square root
of $I_{2} + \frac{1}{2}GG^{T}$ and let $H_{\theta_{0}}$
be the unique positive definite square root of
and $I_{2} + F^{T}_{\theta_{0}}F_{\theta_{0}}$.

\item Then
\[
P = \left (\begin{array}{cc}
E_{\theta_{0}} & F_{\theta_{0}}\\
F_{\theta_{0}}^{T} & H_{\theta_{0}}
\end{array}
\right )
\] is the positive definite factor in the polar decomposition of $X$.
\end{enumerate}
}
\end{algorithm}
     
\begin{remark}
\label{point}
{\rm
\begin{itemize}
\item One could have also obtained the positive definite factor
in the polar decomposition by using the constructive algorithm for
diagonalizing a symmetric $4\times 4$ matrix in \cite{ni}.
This would require solving certain analytic equations for
three variables, whereas in the previous algorithm the number of
unknowns is one, viz., one of the $\theta_{i}$. Furthermore, only when
${\mbox det}(B) = 0$ do these equations require any
work beyond. 

\item The previous algorithm can, with some modifications, be extended
to the Lorentz group (more precisely $G_{I_{1,3}}$). The main difference
is that the $F$ in Step2 would be rectangular, and hence instead
of using $U_{\theta}$ or $V_{\theta}$ one would accordingly
use $3\times 3$ orthogonal
matrices. This would therefore render the corresponding
analytic equations to
involve 3 variables - e.g., the Euler angles of these $3\times 3$ rotations.
We will see however, that passage to its covering group provides
a very tractable alternative to finding the polar decomposition in
the Lorentz group.  
The pros and cons of a similar passage to
the covering group of $SO^{+}(2,2,\mathbf{R})$ is discussed in
Remark (\ref{Whynot}).
\end{itemize}
}
\end{remark}
   
\section{Quaternion Representations of $G_{I_{2,2}}$}

To develop a quaternionic representation of an $X\in G_{I_{2,2}}$, 
we invoke Proposition (\ref{polardecompi}).  

We therefore first obtain a quaternionic representation of 
the positive definite
factor $P$.

\begin{proposition}
{\rm Let $P$ be  positive definite $P$ and belong to $G_{I_{2,2}}$.
Suppose that its NE $2\times 2$ block is $B =
\left (\begin{array}{cc}
a & b\\
c & d
\end{array}
\right )$.
Then it  has quaternionic 
representation $c(1\otimes 1) + p\otimes i + q\otimes j
+ r\otimes k$, where
\begin{itemize}
\item $c = \frac{\alpha_{1}\cos\theta_{1}
+ \beta_{1}\sin\theta_{1} + \gamma_{1}\cos\theta_{1} + \alpha_{2}\cos\theta_{2}
+ \beta_{2}\sin\theta_{2} + \gamma_{2}\cos\theta_{2}}{4}$.
\item $p = \left(\begin{array}{c}
p_{1} \\
\frac{b+c}{2}\\
\frac{d-a}{2}
\end{array}
\right )$, where
\[
p_{1} = \frac{\alpha_{1}\cos\theta_{1}
+ \beta_{1}\sin\theta_{1} + \gamma_{1}\cos\theta_{1} - \alpha_{2}\cos\theta_{2}
- \beta_{2}\sin\theta_{2} - \gamma_{2}\cos\theta_{2}}{4}
\]
\item $q = \left (\begin{array}{c}
\frac{c-b}{2}\\
q_{2}\\
\frac{\alpha_{2}\sin\theta_{2} + \alpha_{1}\sin\theta_{1}}{2}
\end{array}
\right )$
where
\[
q_{2} 
= \frac{\alpha_{1}\cos\theta_{1}
- \beta_{1}\sin\theta_{1} - \gamma_{1}\cos\theta_{1} + \alpha_{2}\cos\theta_{2}
- \beta_{2}\sin\theta_{2} - \gamma_{2}\cos\theta_{2}}{4}
\]
\item $r = \left (\begin{array}{c}
\frac{a+d}{2}\\
\frac{\alpha_{2}\sin\theta_{2} - \alpha_{1}\sin\theta_{1}}{2}\\
r_{3}
\end{array}
\right )$
with
\[
r_{3} =
\frac{\alpha_{1}\cos\theta_{1}
- \beta_{1}\sin\theta_{1} - \gamma_{1}\cos\theta_{1} - \alpha_{2}\cos\theta_{2}
+ \beta_{2}\sin\theta_{2} + \gamma_{2}\cos\theta_{2}}{4}
\]
\end{itemize}
where i) $\alpha_{1} = \sqrt{ 1 + a^{2} + b^{2}}$; ii) $\beta_{1}
= \frac{ac + bd}{\alpha_{1}}$; iii) 
$\gamma_{1} =
[\frac{ (1 + a^{2} + b^{2})(1 + c^{2} + d^{2})
- (ac + bd)^{2}}{\alpha_{1}}]^{1/2}$; iv)  $\alpha_{2}
= \sqrt{ 1 + a^{2} + c^{2}}$; v) $\beta_{1}=
\frac{ab + cd}{\alpha_{2}}$; vi) $\gamma_{2} =
[\frac{ (1 + a^{2} + c^{2})(1 + b^{2} + d^{2})
- (ab + cd)^{2}}{\alpha_{2}}]^{1/2}$; vii) $\tan\theta_{1}
= \frac{\beta_{1}}{\alpha_{1} + \gamma_{1}}$;
viii) $\tan\theta_{2}
= \frac{\beta_{2}}{\alpha_{2} + \gamma_{2}}$;}
\end{proposition}

\noindent {\bf Proof:} In Section 3.1, we saw that any real $B$ can serve
as the NE block of a positive definite $P$ in $G_{I_{2,2}}$.
In this case the NW block has to be the unique positive square root
of $I + BB^{T}$ and the SE block has to be unique positive square root
of $I + B^{T}B$. The result now follows from the formulae for
the positive definite square root in
Lemma (\ref{NoDiagEvenfor2by2}). $\diamondsuit$

Next we address the representation of orthogonal matrices
in $G_{I_{2,2}}$.
 
\begin{proposition}
{\rm Let $X$ be a $4\times 4$ orthogonal matrix which is also in
$G_{I_{2,2}}$. Then 
\begin{enumerate}
\item If $X\in SO^{+}(2,2, \mathbf{R})$, then the quaternionic representation
of it is $u\otimes v$, with $u = a + bi$ and $v= c+ di$, with
$u$ and $v$ both of unit length.
\item If ${\mbox det}(X) = 1$, but $X$ does not belong to
$SO^{+}(2,2, \mathbf{R})$, then the quaternionic representation
of it is $u\otimes v$, with $= \alpha j + \beta k$ and $v =
\gamma j + \delta k$,  with
$u$ and $v$ both of unit length.
\item If ${\mbox det}(X) = -1$, then $X$ has representation
$\frac{1}{2}
(1\otimes 1 + i\otimes i + j\otimes j
- k\otimes k ) (u\otimes v)$, with $u$ and $v$ as in either
 1) or 2) above.
\end{enumerate}
}
\end{proposition}

\noindent {\bf Proof:} We refer to Remark (\ref{Whynot}) to
see that a special orthogonal matrix in $G_{I_{2,2}}$ must be block-diagonal,
with both blocks orthogonal and  both with determinant either $1$ or both
with determinant $-1$. The statement in 1) pertains to the determinant
$1$ case. The statement in 2) pertains to the determinant $1$ case.
Put differently, in the former case $u$ and $v$ both commute with $i$,
while in the latter case they both anticommute with $i$.

The matrix $Y = I_{3,1}$ whose
$2\times 2$ block representation is
${\mbox diag}(I_{2}, \sigma_{z})$,
certainly has negative determinant and is 
in $G_{I_{2,2}}\cap O(4, \mathbf{R})$. So any
orthogonal matrix with determinant $-1$ which is also in
$G_{I_{2,2}}$ must be expressible as $ZY$, where $Z\in
G_{I_{2,2}}\cap SO(4, \mathbf{R})$. Since the quaternionic representation
of $Y$ is precisely $\frac{1}{2}
(1\otimes 1 + i\otimes i + j\otimes j
- k\otimes k )$ the statement in 3) follows. $\diamondsuit$.

Combining the last two results one obtains a quaternionic representation
of any matrix in $G_{I_{2,2}}$. 
\begin{remark}
\label{4porti}
{\rm Find the quaternionic representation of $Q$ from the entries
of $Q$ is susceptible to the possibly folklore method of finding
the pair of unit quaternions from the entries of a given
special orthogonal matrix - see e.g., \cite{FourportI}.
}
\end{remark}

\begin{remark}
\label{FPI}
{\rm The quaternionic representation of matrices in
$Sp(4, \mathbf{R})$ was obtained by starting from the
defining relations for $Sp(4, \mathbf{R})$. If we attempt
to do this for $G_{I_{2,2}}$ then one is lead to
a system of quadratic equations whose structure is not
very transparent. 
 Specifically, let us try to characterize real symmetric
matrices in $G_{I_{2,2}}$. Such a matrix will have quaternionic
representation $a1\otimes 1 + p\otimes i + q\otimes j 
+ r\otimes k$ with \[
a1\otimes 1 + p\otimes i + q\otimes j 
+ r\otimes k (i\otimes i) a1\otimes 1 + p\otimes i + q\otimes j 
+ r\otimes k= i\otimes i
\]

Then a direct, but elaborate, calculation shows that the above equation is
equivalent to the following equations: 
\begin{equation}
\label{One}
ap_{1} = (r\times q)_{1} 
\end{equation}

\begin{equation}
\label{J}
a(r\times i) + q_{1}p + p_{1}q = (p.q)i
\end{equation} 

\begin{equation}
\label{K}
a (i\times q) + r_{1}p + p_{1}r = (p.r)i
\end{equation}

\begin{equation}
\label{IOne}
  a^{2} - \mid\mid p\mid\mid^{2} + \mid\mid q\mid\mid^{2}
+ \mid\mid r\mid\mid^{2} + 2p_{1}^{2} - 2q_{1}^{2} - 2r_{1}^{2} = 1
\end{equation}

\begin{equation}
\label{ITwo}
p_{1}\hat{p} = q_{1}\hat{q} + r_{1}\hat{r}
\end{equation}

\noindent wherein the following {\bf notation} has been eemployed:
Identifying $p, q, r$ with vectors in $\mathbf{R}^{3}$, we write
$\hat{p}$ etc., for their projection on to the $\{j, k\}$ plane.

Equations (\ref{One}) through (\ref{ITwo})
are significantly more complicated to analyse than
those for $Sp(4,\mathbf{R})$.}

\end{remark}

\section{The Lorentz Group and $G_{I_{1,3}}$}
In this section we proceed differently for producing the
polar decomposition and the quaternionic representation of
$G_{I_{1,3}}$. Namely, the fact that the positivity of 
the diagonal blocks being sufficient for positivity of a symmetric
$X\in G_{I_{1,3}}$ is combined with the covering of $SO^{+}(1,3, \mathbf{R})$
by $SL(2, \mathbf{C})$.

Let us recall this covering. Given a $G\in SL(2, \mathbf{C})$,
$\Phi (G)$ is the $4\times 4$ matrix representing the linear
map which sends a $2\times 2$ Hermitian matrix $X$ to $GXG^{*}$
(written with respect to the basis $\{I_{2}, \sigma_{x}, \sigma_{y},
\sigma_{z}\}$). Then $\Phi (G)$ is in $SO^{+}(1,3, \mathbf{R})$.
Furthermore, $\Phi$ is onto and has kernel $\{\pm I_{2}\}$.

We omit the explicit expression of $\Phi (G)$. However, here are
some consequences:
\begin{itemize}
\item i) The $(1,1)$ entry of a matrix in $SO^{+}(1,3, \mathbf{R})$, as is well
known,
is always positive.
Therefore, $-I_{4}$ cannot be in $SO^{+}(1,3, \mathbf{R})$.
Hence $X$ is in $SO^{+}(1,3, \mathbf{R})$, iff $-X$ is in
the other connected component of the determinant one matrices
in $G_{I_{1,3}}$.
\item $\Phi (G^{*}) = [\Phi (G)]^{T}$. This follows from the explicit
expression for $\Phi (G)$, but can also be shown by calculating the
adjoint of the linear map $X\rightarrow GXG^{*}$.
\end{itemize}

We can now characterize the symmetric matrices and also the positive
definite matrices in $SO^{+}(1,3, \mathbf{R})$.

\begin{theorem}
\label{PosDefInLorenz}
{\rm A symmetric matrix in $SO^{+}(1,3, \mathbf{R})$ is either
positive definite or indefinite. The former are $\Phi$ images of
Hermitian matrices in $SL(2, \mathbf{C})$ and the latter are
$\Phi$ images of anti-Hermitian matrices in $SL(2, \mathbf{C})$.
Furthermore, a positive definite matrix in
$SO^{+}(1,3, \mathbf{R})$  is always expressible as $\Phi (G)$ for
a positive definite $G\in SL(2, \mathbf{C})$.}
\end{theorem}

{\bf Proof:} Every matrix in $SO^{+}(1,3, \mathbf{R})$ is
a $\Phi (G)$ for some $G\in SL(2, \mathbf{C})$. Let $\Phi (G)$ be symmetric.
Hence, $\Phi (G^{*}) = \Phi (G)$ be virtue of the discussion above.
Hence $G^{*} = \pm G$. Let $G^{*} = G$ first. So
$G = \left (\begin{array}{cc} a & x + iy\\
x-iy & d
\end{array}
\right )$. Now a calculation of $\Phi (G)$, wherein repeated use
of the fact that $ad - (x^{2} + y^{2}) = 1$ is made, reveals
it to be
\[
\Phi (G) = \left (\begin{array}{cccc}
\frac{(a+d)^{2}}{2} - 1 & (a+d)x & -(a+d)y & \frac{a^{2}-d^{2}}{2}\\
(a+d)x & 1 + 2x^{2} & -2xy & (a-d)x\\
-y(a+d) & -2xy & 1 + 2y^{2} & y(d-a)\\
\frac{a^{2}-d^{2}}{2} & (a-d)x & y(d-a) & \frac{(a-d)^{2}}{2} + 1    
 \end{array}
\right )
\]
To show that it is positive we just need to verify that the diagonal
blocks are positive definite, thanks to  1) of Remark(\ref{LorenzToo}).
Now the $(1,1)$ entry is positive (because it equals half of $\mid\mid G\mid
\mid_{F}^{2}$). Next the determinant of the NW block
is $\frac{(a+d)^{2}}{2} -1 - 2x^{2}$. Using again the fact that
$ad - (x^{2} + y^{2}) = 1$, this equals $\frac{a^{2} + d^{2} + 2y^{2}
- 2x^{2}}{2}$. Next $a^{2} + d^{2} \geq 2ad = 2 (1 + x^{2} + y^{2})$.
So $a^{2} + d^{2} + 2y^{2}
- 2x^{2} \geq 2 + 4y^{2}$. Hence the determinant of the NW block is
positive. The $(3,3)$ entry is visibly positive, while the determinant
of the SE block equals $ 1 + 2y^{3} + \frac{(a-d)^{2}}{2}$ which
is, of course, positive. Hence $\Phi (G)$ is positive definite.
Now a Hermitian $G$ in $SL(2, \mathbf{C})$ is obviously either positive
definite or negative definite. Hence, if we show that the $\Phi (G)$ image of
an anti-Hermitian matrix cannot be positive definite, then it will follow
that $X\in SO^{+}(1,3, \mathbf{R})$ is positive definite iff it is expressible
as $\Phi (G)$ for a positive definite $G$.

Now consider an anti-Hermitian $G = \left (\begin{array}{cc}
ia & x + iy\\
-x + iy & id
\end{array} \right )$. We wish to show $\Phi (G)$ is indefinite. Since the
$(1,1)$ entry is positive for any $\Phi (G)$, it cannot be negative definite.
So it suffices to show that is not positive definite. To this end
we record the NW and SE blocks of $\Phi (G)$, making use of
$x^{2} + y^{2} - ad = 1$:
\[
{\mbox NW} = \left (\begin{array}{cc}
\frac{(a+d)^{2}}{2} + 1 & (a+d)y\\
(a+d)y & -1 + 2y^{2}
\end{array}\right )
\]

\[
{\mbox SE}
 = \left (\begin{array}{cc}
- 1 + 2x^{2} & x(a-d)\\
x(a-d) & \frac{(a-d)^{2}}{2} - 1
\end{array} \right )
\]

We will now show that at least one of the 
determinants of these two blocks most be non-positive.
Suppose, to the contrary, that both are positive. Then
$-1 - \frac{(a+d)^{2}}{2} + 2y^{2} > 0$ and $1- \frac{(a-d)^{2}}{2} - 2x^{2}
> 0$. Subtracting the second inequality from the first one obtains
$-2ad + 2y^{2} + 2x^{2} > 2$, which contradicts
$-ad + x^{2} + y^{2} = 1$.

So $\Phi (G)$ is indefinite. $\diamondsuit$

The above proof yields the following 
algorithm to find a positive definite $G\in  
SL(2, \mathbf{C})$ satisfying $\Phi (G) = P$ for a positive definite
$P\in SO^{+}(1,3, \mathbf{R})$ essentially by inspection, to wit:

\begin{algorithm}
\label{PosDefInversioninLorentz}
Inversion of $\Phi: SL(2, \mathbf{C}\rightarrow SO^{+}(1,3, \mathbf{R})$,
when the target is a positive definite P:

{\rm \begin{enumerate}
\item First suppose $P_{22}\neq 1$ and $P_{33}\neq 1$. Then let $x$
be one of the two square roots of $\frac{P_{22}-1}{2}$ and $y$
one of the two square roots $\frac{P_{33}-1}{2}$. Corresponding to
each such choice (there are four of them) solve the equations
$a + d = -\frac{P_{13}}{y}$ and $a-d = \frac{P_{24}}{x}$. Precisely
one of these pairs will have a solution $(a, d)$ with both $a$ and $d$
positive.  Choose the corresponding $x, y, a, d$. Then $H
= \left ( \begin{array}{cc}
a & x + iy\\
x-iy & d
\end{array}
\right )$ satisfies $\Phi (H) = G$.
\item If $P_{12} = 1$ and $P_{33}\neq 1$, pick $x=0$ and $y$ to
be one of the square roots of $\frac{P_{33}-1}{2}$. Corresponding to
each choice of $y$, solve the system $ a + d = -\frac{P_{31}}{y}, d-a
= \frac{P_{34}}{y}$. Precisely one of these will have a solution
with both $a$ and $d$ positive. Choose the corresponding $y,a,d$ and
let $H
= \left ( \begin{array}{cc}
a & iy\\
-iy & d
\end{array}
\right )$. Then $\Phi (H) = G$   

\item If $P_{12} \neq 1$ but $P_{33} = 1$, then set $y=0$ and let
$x$ be one of the two square roots of
$\frac{P_{22}-1}{2}$. Corresponding to each choice of $x$ solve the
system $a + d = \frac{P_{12}}{x}, a-d = \frac{P_{24}}{x}$. Precisely
one of these will have solution with both $a$ and $d$ positive. Choose
this value of $x$ and let $H
= \left ( \begin{array}{cc}
a & x\\
x & d
\end{array}
\right )$. Then $\Phi (H) = G$.
\item Finally, if both $P_{22} =1$ and $P_{33} =1$, then set
$x=y = 0$, and let $\alpha$ be one of the two square roots
of $2(P_{11} + 1)$ and $\beta$ one of the two square roots of
$2(P_{44} -1)$. For each choice solve $a+ d = \alpha ; a-d = \beta$.
Precisely one of these will have a solution with both $a$ and $d$ positive.
Let $H
= \left ( \begin{array}{cc}
a & 0\\
0 & d
\end{array}
\right )$. Then $\Phi (H) = G$.
\end{enumerate}
}
\end{algorithm}  

With the algorithm above, we can now constructively
find the polar decomposition of a matrix in the 
polar decompostion of any matrix in the Lorentz group and hence in
all of $G_{I_{1,3}}$.

First consider a matrix $X\in SO^{+}(1, 3, \mathbf{R})$. Then $X^{T}X$ is also
in $X\in SO(1, 3, \mathbf{R})$ and positive definite. Then one finds a $G\in
SL(2, \mathbf{C})$, with $\Phi (G) = X^{T}X$. Let $H\in SL(2, \mathbf{C})$ be
the unique positive definite square root of $G$. Thus $H^{2} = G$,
and hence $(\Phi (H))^{2} = X^{T}X$. By Theorem (\ref{PosDefInLorenz}),
$\Phi (H)$ is positive definite and hence must be the positive
part $P$ of the polar decomposition $X = VP$ of $X$. Now finding
$V$ is routine. Furthermore, since $V$ is also in
$SO^{+}(1, 3, \mathbf{R})$, it must be special orthogonal and also
satisfy $VI_{1,3} = I_{1,3}V$. Therefore, its first column and row must
be the first standard unit vector. Hence $V$ is an (standard) embedding of a 
matrix in $SO(3, \mathbf{R})$ into a $4\times 4$ matrix. Thus $V$ is
represented by $v\otimes v$ for some unit quaternion $v$.

Next, let $X\in SO(1, 3, \mathbf{R})$ have determinant $1$ and a negative
$(1,1)$ entry. Hence $-X \in SO^{+}(1, 3, \mathbf{R})$. Computing
$-X = VP$, we find $X = (-V)P$ is the polar decomposition of $-X$. Thus,
in particular the orthogonal factor has quaternionic representation
$-u\otimes u$ for some unit quaternion $u$.

Suppose $X\in G_{I_{1,3}}$ has determinant equal to $-1$ and its
$(1,1)$ entry is positive.Then $X = I_{1,3}Y$, where 
$Y\in SO^{+}(1, 3, \mathbf{R})$. Clearly, $Y^{T}Y = X^{T}X$ and thus
the positive definite factor in $X$'s polar decomposition coincides
with that of $Y$, while its orthogonal factor is clearly $I_{1,3}V$,
where $V$ is that of $Y$. Hence its quaternionic representation is
$\frac{1}{2}
(1\otimes 1 - i\otimes i - j\otimes j
- k\otimes k ) v\otimes v$, for a unit quaternion $v$.

Finally, let $X\in G_{I_{1,3}}$ has determinant equal to $-1$ and suppose its
$(1,1)$ entry is negative. Then $-X = I_{1,3}Y$, with
$Y\in SO^{+}(1, 3, \mathbf{R})$. Once again the positive definite factor
of $X$ coincides with that of $Y$, and its orthogonal factor is
$-I_{1,3}V$, where $V$ is that of $Y$. So its quaternionic
representation is $-\frac{1}{2}
(1\otimes 1 - i\otimes i - j\otimes j
- k\otimes k ) v\otimes v$, for a uniq quaternion $v$.

The foregoing arguments are now summarized as an algorithm:

\begin{algorithm}
\label{PolarinLorenz}
{\rm \begin{enumerate}
\item Let $X_{11} > 0$ and ${\mbox det}(X) = 1$. Compute $X^{T}X$.
Use Algorithm (\ref{PosDefInversioninLorentz}) to find the positive definite
$G\in SL(2, \mathbf{C})$ satisfying $\Phi (G) = X^{T}X$.
\item Compute the unique positive definite square root 
$H\in SL(2, \mathbf{C})$ of $G$. Find $P= \Phi (H)\in SO^{+}(2,2,\mathbf{R})$.
Then $P$ is the positive definite factor in the polar decomposition
of $X = VP$. The orthogonal factor $V$ equals $I_{1,3}PI_{1,3}$.
Since $I_{1,3}$ is diagonal, this computation requires only $2\times 2$
matrix multiplications.
\item Next let $X_{11} < 0$ and ${\mbox det}(X) = 1$. Then $-X\in
SO^{+}(2,2,\mathbf{R})$. Apply Steps 1 and 2 to $-X$ to find the
polar decomposition $-X = VP$. Then $X = (-V)P$ is the polar decomposition
of $X$.
\item Finally let ${\mbox det}(X) = -1$. Then 
$I_{1,3}X\in SO(2,2, \mathbf{R})$. Apply the previous steps to
find the polar decomposition $I_{1,3}X = VP$. Then $X = I_{1,3}VP$
is the polar decomposition fo $X$.
\end{enumerate}
}
\end{algorithm}    

Next we find the quaternionic representation of any matrix
in $G_{I_{1,3}}$:

\begin{theorem}
\label{HRepforLorenz}
{\rm Let $X\in G_{I_{1,3}}$.
Then
\begin{enumerate}
\item If $X_{11} > 0$ and ${\mbox det}(X) = 1$, then
$X$'s quaternionic representation is
$ [u\otimes u][c1\otimes 1 + p\otimes i + q\otimes j + r\otimes k]$,
where $u$ is a unit quaternion and $p, q, r$ and $c$ are as follows: 

\[
p = \left [\begin{array}{c}
x^{2} \\
(a^{2} - d^{2} - 4xy)/4\\
( (a-d)x + (a+d)y)/2
\end{array}
\right ]
\]

\[
q = \left [\begin{array}{c}
(d^{2} - a^{2} -4xy )/4\\
y^{2}\\
( (a + d)x + (d-a)y)/2
\end{array}
\right ]
\]

\[
r = \left [\begin{array}{c}
((a-d)x - (a+d)y)/2\\
(y (d-a) - (a+d)x)/2\\
(a-d )^{2}/4
\end{array}
\right ]
\]

\[
c =  (a+d)^{2}/4
\]

with $ad - (x^{2} + y^{2}) = 1$ and $a>0$.

\item If $X_{11} < 0$ and ${\mbox det}(X) = 1$, then the representation
of $X$ is $ -[u\otimes u][c1\otimes 1 + p\otimes i + q\otimes j + r\otimes k]$ 
with $p, q, r, c$ and $u$ as in Case 1. 
\item If $X_{11} > 0$ and ${\mbox det}(X) = -1$, then the representation
is  $ \frac{1}{2}(1\otimes 1 - i\otimes i
- j\otimes j - k\otimes k)[u\otimes u]
[c1\otimes 1 + p\otimes i + q\otimes j + r\otimes k]$,
with $p, q, r, c$ and $u$ as in Case 1.

\item If $X_{11} < 0$ and ${\mbox det}(X) = -1$, then the representation
is  $ -\frac{1}{2}(1\otimes 1 - i\otimes i
- j\otimes j - k\otimes k)[u\otimes u]
[c1\otimes 1 + p\otimes i + q\otimes j + r\otimes k]$,
with $p, q, r, c$ and $u$ as in Case 1.
 
\end{enumerate}
}
\end{theorem}
\begin{remark}
\label{Whynot}
{\rm A natural question that arises is if one could not have profitably
used the covering $SL(2, \mathbf{R})\times SL(2, \mathbf{R})
\rightarrow SO^{+}(2,2,\mathbf{R})$ to compute the polar decomposition
in $SO^{+}(2,2,\mathbf{R})$, in a manner analogous to that for the
Lorentz group. 
The principal obstruction is that the
inversion of the covering map when the target in $SO^{+}(2,2,\mathbf{R})$
is a positive definite matrix is no longer any simpler than it is
for a general matrix in $SO^{+}(2,2,\mathbf{R})$, in sharp contrast
to the Lorentz case, \cite{francis}. Nevertheless this exercise
reveals the following facts of independent interest:

\begin{enumerate}
\item $\Phi (A^{T}, B^{T}) = [\Phi (A, B)]^{T}$ as before.
\item If $A, B\in SL(2, \mathbf{R})$ are both positive definite or
both negative definite, then $\Phi (A, B)$ is positive definite and
only then is a symmetric matrix in $SO^{+}(2,2,\mathbf{R})$ positive definite.
\item Unlike the case for the Lorentz group, there are negative definite
matrices in $SO^{+}(2,2,\mathbf{R})$ and these are precisely the
image of a pair $(A, B)$ where one is positive definite and the
other negative definite.
\item If both $A$ and $B$ are antisymmetric, then $\Phi (A, B)$ is either
$I_{2,2}$ or $-I_{2,2}$. Therefore, unlike the Lorentz group,
$-I_{4}\in SO^{+}(2,2,\mathbf{R})$.
\item A determinant one matrix in $G_{I_{2,2}}$ is special orthogonal
iff it is block-diagonal. This is because these two conditions
imply that $XI_{2,2} = I_{2,2}X$. Obviously both blocks have to be
orthogonal. Such a matrix is in $SO^{+}(2,2, \mathbf{R})$
iff both these diagonal blocks are in $SO(2, \mathbf{R})$.
If both blocks have determinant equal to $-1$,
then the matrix cannot be in $SO^{+}(2,2,\mathbf{R})$. This follows
from the explicit form of the covering homomorphism
$\Phi: SL(2, \mathbf{R})\times SL(2, \mathbf{R})
\rightarrow SO^{+}(1,2,\mathbf{R})$. Indeed, the only solutions $(A, B)$
to $\Phi (A, B) = X$, with $X$ block-diagonal, and both blocks
orthogonal and
with determinant equal to $-1$ turn out to be complex. The
same reasoning extends to $SO(n,n,\mathbf{R})$. Specifically,
the map which assigns to a matrix $X$ the determinants of its
diagonal blocks is continuous. For $X\in SO(n,n,\mathbf{R})\cap
SO(2n, \mathbf{R})$ this function assumes only the values
$(1,1)$ and $(-1,-1)$. Therefore, the these determine the
to connected components of $SO(n,n,\mathbf{R})\cap
SO(2n, \mathbf{R})$. 

This circumstance
therefore distinguishes the two connected components of 
$\{X: X\in G_{I_{n,n}}, {\mbox det}(X) =1\}$ - the orthogonal factor
in the polar decomposition of $SO^{+}(n,n,\mathbf{R})$ is block-diagonal
with both blocks in $SO(n,\mathbf{R})$, whereas the orthogonal factor
in the polar decomposition of matrices in the other connected component
have both diagonal blocks orthogonal with determinant $-1$.
\end{enumerate}

We will content ourselves
with the proof of the fact that if $A$ and $B$ are positive definite
then so is $\Phi (A, B)$. To that end we let $A = \left (\begin{array}{cc}
x_{1} & x_{2}\\
x_{2} & x_{8}
\end{array}
\right )$ and $B = \left (\begin{array}{cc}
x_{3} & x_{4}\\
x_{4} & x_{6}
\end{array}
\right )$ 
be a pair of positive definite matrices in $SL(2, \mathbf{R})$.
Following \cite{fpi} we embed this pair concentrically in $M(4, \mathbf{R})$
as follows:

\[
C = \left (\begin{array}{cccc}
x_{1} & 0 & 0 & x_{2}\\
0 & x_{3} & x_{4} & 0\\
0 & x_{5} & x_{6} & 0\\
x_{7} & 0 & 0 & x_{8}
\end{array}
\right )
\]

(This unusual embedding is what naturally results from the 
iterative constructions in Clifford algebras, \cite{fpi}.) 

Then the covering map is the map which sends $(A, B)$ to the 
matrix of the linear map $X \rightarrow CXC^{-1}$, with respect to the
basis $\{ \sigma_{z}\otimes \sigma_{x}, \sigma_{x}\otimes I_{2},
\sigma_{z}\otimes i\sigma_{y}, i\sigma_{y}\otimes I_{2}\}$.  
We don't need the full matrix to prove our claim, since a
matrix in $SO^{+}(2,2, \mathbf{R})$ is positive definite iff its diagonal
blocks are. Accordingly we display only these blocks. They are
\[
{\mbox NW} = \frac{1}{2}
\left (\begin{array}{cc}
x_{1}x_{6} + x_{3}x_{8} + 2x_{2}x_{4} & 2x_{2}x_{4}\\
2x_{2}x_{4} & x_{1}x_{3} + x_{6}x_{8} -2x_{2}x_{4}
\end{array}
\right )
\]

and
\[
{\mbox SE} = \frac{1}{2}
\left (\begin{array}{cc}
x_{1}x_{6} + x_{3}x_{8} - 2x_{2}x_{4} & 2x_{2}x_{4}\\
2x_{2}x_{4} & x_{1}x_{3} + x_{6}x_{8} + 2x_{2}x_{4}
\end{array}
\right )
\]

Now $x_{1}x_{6} + x_{3}x_{8}\geq 2 \sqrt{x_{1}x_{6}x_{3}x_{8}}
= 2\sqrt{x_{1}x_{8}}\sqrt{x_{3}x_{6}} > 2\sqrt{x_{2}^{2}}\sqrt{x_{4}^{2}}
= 2\mid x_{2}x_{4}\mid$
which shows that $d_{11}$ and $d_{33}$ are positive.

The determinant of the NW 
block equals $x_{1}x_{8}x_{3}x_{4} - x_{2}^{2}x_{4}^{2}$, after a calculation 
which uses
$1 - x_{1}x_{8} = -x_{2}^{2}$ and $1-x_{3}x_{6} = -x_{4}^{2}$. 
Obviously this is positive since $x_{1}x_{8} > x_{2}^{2}$ and
$x_{3}x_{6} > x_{4}^{2}$. One similarly shows that the determinant of
the SE block is also positive. Thus, $\Phi (A, B)$ is positive definite. 
}
\end{remark}

The following result which stems from Remark (\ref{Whynot}) is worth
recording separately:

\begin{theorem}
\label{CCCharacterize}
{\rm Let $X\in G_{I_{n,n}}$ with its diagonal
$n\times n$ blocks denoted $A$ and $D$.
The four connected components of $G_{I_{n,n}}$ admit the
following description:
\begin{itemize}
\item ${\mbox det}(X) = 1, {\mbox det}(A) > 0$ and ${\mbox det}(D) > 0$.
This is $SO^{+}(n,n,\mathbf{R})$.
\item ${\mbox det}(X) = 1, {\mbox det}(A) < 0$ and ${\mbox det}(D) < 0$.

\item ${\mbox det}(X) = -1, {\mbox det}(A) > 0$ and ${\mbox det}(D) < 0$.

\item ${\mbox det}(X) = -1, {\mbox det}(A) < 0$ and ${\mbox det}(D) > 0$.
\end{itemize}
}
\end{theorem}

\noindent {\bf Proof:} Let $X = VP$ be the polar decomposition of $X$.
By Remark (\ref{LogofPosDef}), $P$ always belongs to $SO^{+}(n,n,\mathbf{R})$
and thus since $SO^{+}(n,n,\mathbf{R})$ is a group, we see that if $X\in
SO^{+}(n,n,\mathbf{R})$, then so is $V$. Hence by 5) of Remark
(\ref{Whynot}), $V$ is block-diagonal with both blocks having positive
determinant. Since both diagonal blocks of $P$ have positive determinants,
it follows that the same holds for $X$. Next let ${\mbox det}(X) = 1$,
but let $X$ not belong to $SO^{+}(n,n,\mathbf{R})$. 
If $X = VP$, then it follows that $V$ cannot be in $SO^{+}(n,n,\mathbf{R})$,
and hence by 5) of Remark (\ref{Whynot}), $V$ is block-diagonal
with both blocks having negative determinants. Hence, the same holds
for $X$. Next, let ${\mbox det}(X) = -1$. Then obviously,
$X = {\mbox diag}(I_{n},I_{n,1})Y$, with $Y\in SO(n,n,\mathbf{R})$.
Hence the remaining statements follow. $\diamondsuit$. 
\section{Conclusions}
In this work algorithms for computing the entries of
the matrices in a polar decomposition of matrices preserving
the non-degenerate bilinear form with signature matrices $I_{2,2}$ and
$I_{1,3}$, as well as their quaternionic representations
were provided. These methods eschewed any g
eigencalculations.
In the process an interesting characterization of positive definite
matrices and methods for completion to positivity
in these groups were obtained. In addition, an explicit formula for
``the" logarithm of a positive definite matrix in $G_{I_{n,n}}$ was
obtained. 

One  question suggested by this work that seems worth
pondering over is the following. Is it true, in general,   
that the preimage of a positive definite matrix in $SO^{+}(p,q,\mathbf{R})$
in its covering group can also be chosen to be positive definite?

\section{Appendix}
In this appendix the quaternionic characterization of symmetric and
positive definite matrices in $Sp(4, \mathbf{R})$ is provided for
purposes of comparison.

\begin{theorem}
\label{symsymplectic}
{\rm Let $X$ be a $4\times 4$ symplectic matrix which is also symmetric.
	Then it admits the quaternion representation
		$X = a1\otimes 1 + p\otimes i + q\otimes j + r\otimes k$, with
$aq =r\times p$, $p.q = 0 = r.q$, and $a$ satisfying the constraint
$a^{2}-p.p + q.q - r.r = 1$. If $a = \frac{1}{4} {\mbox Tr}(X)\neq 0$,  
then $X$ is symplectic iff 
$aq = r\times p$ and $a^{2}-p.p + q.q - r.r = 1$.
Such an $X$ is positive definite in addition, iff i)$a >0$ and ii)
$2a^{2} -2(q.q) + 1 > 0$. In particular, a symmetric, symplectic
matrix with $a > 0, q = 0$ is always positive definite.  
}
\end{theorem}

It is primarily the fact that the vector $q$ is essentially the cross-product
of $p$ and $r$ that made the above quaternionic representation as the
most convenient starting point for the algorithmic determination of
the polar decomposition in the symplectic group in \cite{noncompactportion}.

\section{Acknowledgements}
This work was supported by NSF's Louis Stokes Alliance for
Minority Participants (LSAMP) program and the Clark Scholars program
at the University of Texas at Dallas.
\section{Competing Interests}
On behalf of each author, the corresponding author states that there
is no conflict of interest.

\end{document}